\documentclass[%
 reprint,
superscriptaddress,
 amsmath,amssymb,aps, prl, floatfix, showpacs,citeautoscript
]{revtex4-1}

\usepackage{graphicx,graphics}
\usepackage{dcolumn}
\usepackage{bm}
\usepackage{amsfonts,amssymb,amsmath}
\usepackage[colorlinks,linkcolor={blue},citecolor={blue},urlcolor={blue}]{hyperref}
\usepackage{siunitx}
\usepackage{textcomp}
\usepackage{float}

\usepackage{color}
\usepackage{array}
\usepackage{booktabs}
\usepackage{multirow}
\usepackage{color}
\usepackage{epsfig}\usepackage{subfigure}
\usepackage{dcolumn}
\usepackage{graphicx,graphics}
\usepackage{amsmath}
\usepackage{amssymb}
\usepackage{bm}
\usepackage{color}
\usepackage[utf8]{inputenc}
\begin{document}

\title{Proximity-induced topological phases in bilayer graphene}

\author{Abdulrhman M. Alsharari}
\email{aalsharari@ut.edu.sa}
\affiliation{Department of Physics and Astronomy, and Nanoscale and Quantum
	Phenomena Institute, \\ Ohio University, Athens, Ohio 45701}
\affiliation{Department of Physics, University of Tabuk, Tabuk, 71491, SA}
\author{Mahmoud M. Asmar}
\affiliation{Department of Physics and Astronomy, Louisiana State University, Baton Rouge, LA 70803}
\author{Sergio E. Ulloa}
\affiliation{Department of Physics and Astronomy, and Nanoscale and Quantum
	Phenomena Institute, \\ Ohio University, Athens, Ohio 45701}

\date{\today}

\begin{abstract}
We study the band structure of phases induced by depositing bilayer graphene on a transition metal dichalcogenide monolayer.  Tight-binding and low-energy effective Hamiltonian calculations show that it is possible to induce topologically nontrivial phases that should exhibit spin Hall effect in these systems.  We classify bulk insulating phases through calculation of the Z$_2$ invariant, which unequivocally identifies the topology of the structure.  The study of these and similar hybrid systems under applied gate voltage opens the possibility for tunable topological structures in real experimental systems.
\end{abstract}

\maketitle


Graphene is a single sheet of carbon atoms characterized by its massless low-energy excitations at opposite corners of the Brillouin zone \cite{graphene}. The existence of such Dirac-like particles at the $K$ and $K'$ valleys is guaranteed by the conservation of both time reversal and inversion symmetry of the lattice.
That makes these points susceptible to symmetry breaking perturbations.

 Several methods relying on adatom deposition have been proposed to diversify graphene's functionalities. Some of these focus on the enhancement of spin-orbit (SO) active perturbations \cite{QAH-GwR&M,theory,Graphene-Ni}, and aim to the realization of the quantum spin Hall effect \cite{qshe}. The adatom-based enhancement of SO also produces a variety of valley and pseudospin dependent perturbations. These additional effects may mask the modification of desired SO couplings \cite{mimimal,mahmoud-PRB}, motivating the search for new systems with a higher degree of control over SO enhancement.

Layered systems involving several graphene layers or hybrids with other two dimensional materials, such as transition metal dichalcogenides (TMDs), have provided new venues towards graphene functionalization. The earliest example of these multilayered systems is bilayer graphene (BLG)  \cite{bilayer}. BLG is characterized by quadratically dispersing bands, degenerate at the $K$ and $K'$ points. This degeneracy is also protected by both time reversal and inversion symmetry. As such, an external gate voltage that reduces the $z\rightarrow -z$ symmetry leads to the generation of a gap
and interesting curvature near the $K$ points in BLG \cite{TI-Bi-G,R-V-Bi-G}. The enhancement of symmetry-allowed SO interactions (Rashba and intrinsic SO) remains negligible, however, due to the light atomic weight of its constituent carbon atoms.

The atomic components in TMDs are heavier and their intrinsically asymmetric lattice structure lead to sizable SO effects \cite{mos24,3B-TB-TMD}. Although SO in TMDs allow for valley and spin selectivity in photoemission experiments \cite{selctive,selctive1,selctive2}, the topological nature of these materials remains trivial \cite{Inv-G-TMDs,Frank-GT}. The gap generated in TMDs is in large part due to the breaking of in-plane inversion symmetry, as well as to the SO interactions that lift the spin degeneracy.
The stacking of the substrate sensitive graphene and SO-active TMD monolayers leads to a hybridization of the low energy states of the resulting heterojunction and a concomitant band inversion under suitable conditions   \cite{G-WS2-Edge_SOC,mogrf2,Inv-G-TMDs}. The hybridization produces systems that include  structures with no band inversion and systems with inverted bands, as different TMDs or gate voltages are used \cite{Inv-G-TMDs}.
The band-inverted regime in this heterolayer system is accompanied by unique one-dimensional metallic edge states, strongly localized near the borders of the heterojunction  \cite{G-WS2-Edge_SOC,mogrf2,Inv-G-TMDs}.

The sensitivity of the charge carrier dispersion in BLG to external electric fields across layers, combined with the SO-active TMD materials, suggests an interesting new approach for the possible manipulation of topological properties in two-dimensional materials. Although experimental findings have been reported on exciton effects \cite{G-WS2-Exciton}, and transport \cite{SOC-G-TMDs,G-WS2-Edge_SOC}, as well as first principles calculations studying the effects of an electric field in these systems \cite{Fabian-BLG-TMD}, the study of possible topological phases has not received much attention. Our work complements theoretical models where BLG is studied in the presence of intrinsic and Rashba SO  \cite{TI-Bi-G,R-V-Bi-G}, with a realistic multilayered system containing BLG and TMD. As we will see, this multilayer exhibits an interesting phase diagram with nontrivial topological phases that should be accessible in the laboratory.

In order to study the BLG-TMD heterostructure, we make use of a multi-orbital tight-binding formalism, complemented by an analysis to determine the symmetry-allowed perturbations in the combined system. The proximity of a TMD to BLG leads to the enhancement of different types of SO coupling in addition to a `staggered' potential that breaks sublattice symmetry. The largest SO contributions are found to be Rashba and Zeeman-like SO terms, much larger than the staggered potential and the intrinsic SO \cite{qshe}.

The BLG-TMD system displays unique electronic and spin properties which reflect its reduced symmetries. This is accompanied by the appearance of band gaps near the $K,K'$ valleys, and the relatively large SO make this system exhibit different topological phases.
We identify the insulating phases and characterize their topology by calculating the corresponding Z$_{2}$ invariants. The Z$_{2}$ topological index reveals nontrivial phases for sufficiently large Rashba SO coupling on both layers. However, we show that as an asymmetric voltage increases, the system undergoes a phase transition to a trivial insulator, even for large values of Rashba couplings.


We adopt a tight-binding model that couples nearest neighbors in graphene and up to
third-nearest-neighbors of the metal atoms, with three $d$-orbitals: $ d_{z^2} $, $ d_{x^2-y^2} $ and $ d_{xy} $.
The Hamiltonian describing the trilayer system can be written as
$
H=H_{{\rm BLG}}+H_{{\rm TMD}} +H_{{\rm tunn}}\;,
$
where $H_{{\rm BLG}}$ is the Hamiltonian of a Bernal-stacked BLG with direct coupling between A and B atoms in corresponding layers \cite{graphene}. $H_{{\rm TMD}}$ describes the TMD monolayer and $H_{{\rm tunn}}$ the coupling between the components of the system.  Due to the incommensurability between graphene and TMD lattices, we analyze heterostructure supercells with different size ratios. We focus here on the 4$ \times $4 TMD with 5$ \times $5 graphene cell--see supplement for details \cite{supp}.

The coupling between layers is parameterized via Slater-Koster couplings between the graphene $ p_z $ and the $d$-orbitals in TMD.
Although the $z\rightarrow -z$ symmetry is intrinsically broken by the stacking geometry, this effect can be further enhanced by applying an electric field that sets the graphene layers at different voltage potentials. The intrinsic and field-induced asymmetry generates  Rashba couplings that mix spin states and plays an important role in the heterostructure properties  \cite{Z2-QSH,Bi-G-TI}.

\begin{figure}[h]
	\centering
    \includegraphics[width=1.0\linewidth]{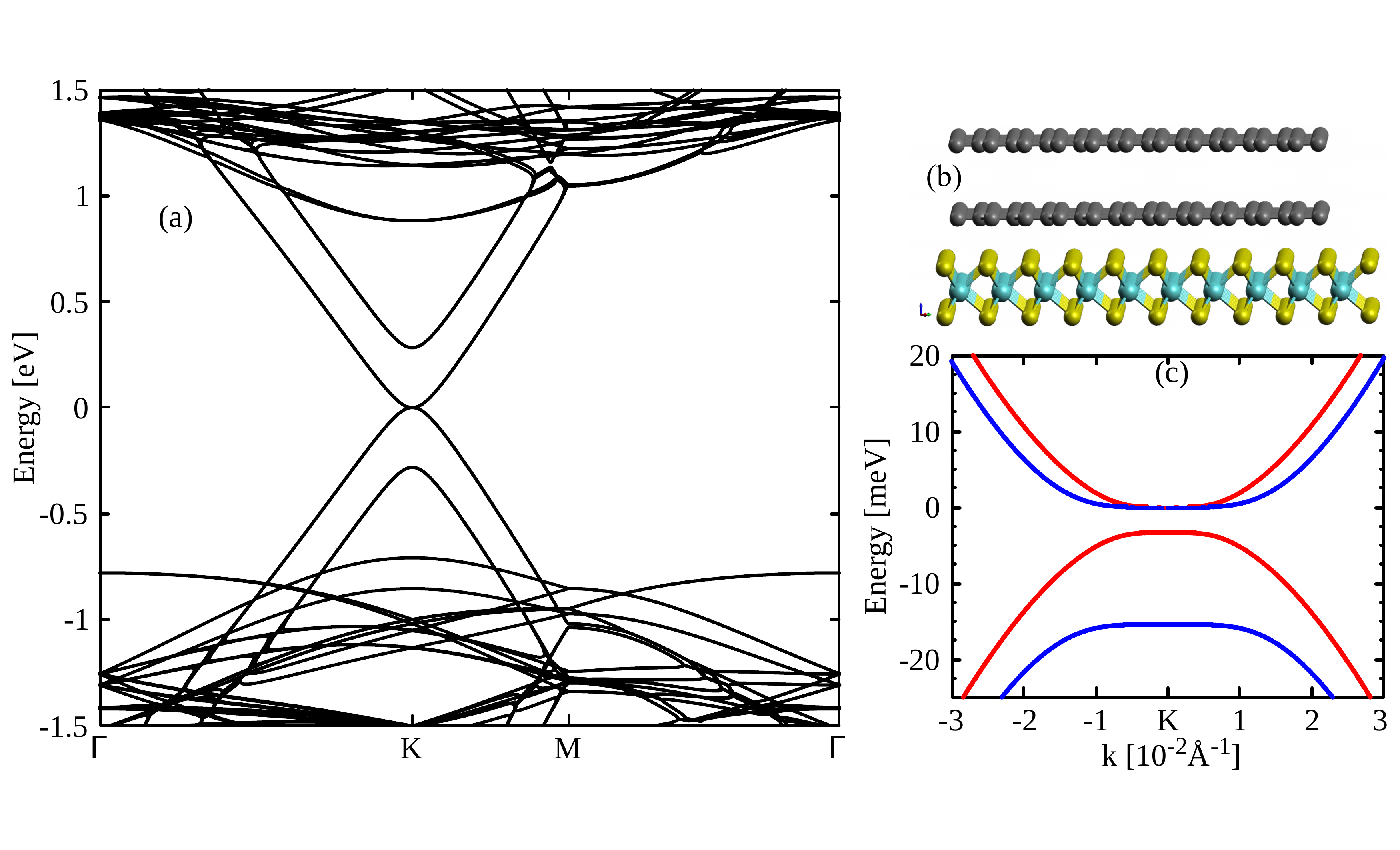}
	\caption{(a) $\Gamma$-$K$-$M$-$\Gamma$ band structure of coupled bilayer graphene-TMD from tight-binding calculations where TMD is WS$_2$.
(b) Side view of atomic arrangement for BLG-TMD system.
(c) Bands at $K$ valley near the Fermi level ($E_{F}=0$). Red (blue) lines indicate $s_{z}$ spin up (down) projection of each band.  }
	\label{Fig1}
\end{figure}

We proceed to analyze the band structure and the effects of externally applied voltages.
 The system is depicted in Fig.\ \ref{Fig1}b. Its reduced in-plane mirror and $z\rightarrow -z$ symmetries are reflected in the gap generated in the electronic dispersion, as seen in Fig.\ \ref{Fig1}c, a zoom-in of Fig.\ \ref{Fig1}a near the Fermi level (here set at $E_F=0$). The low energy structure is dominated by the Bernal interlayer coupling ($\simeq 0.3$ eV \cite{Bilayer-Graphene-TB}), two orders of magnitude larger than the proximity-induced parameters, such as SO couplings and staggered potential shifts.

 The low-energy dispersion at $K$ and $K'$ valleys are similar and host Kramer pair states. The upper and lower bands (with extrema at $\simeq \pm 0.3$ eV) are spin resolved (with splitting not visible in Fig.\ \ref{Fig1}a), while preserving a nearly parabolic shape. However, the bands closer to the Fermi energy exhibit a spectral gap of the order of few ($\simeq 10$) meV
due to the presence of the TMD layer, and acquire a rather non-parabolic shape, Fig.\ \ref{Fig1}c.
These low energy bands exhibit then a massive Dirac structure, similar to that of the parent TMD but with rescaled parameters--details are given in the supplement \cite{supp}.  Notice that the lowest conduction bands at each valley are nearly spin degenerate, lifted by a weak Rashba SO term. However, the spin splitting is dominant in the valence bands, and it is accompanied by weak mass inversion.  The details depend on which TMD is used and on supercell sizes, but the general features remain \cite{supp}.

\begin{figure}[h]
   	\centering
   	\includegraphics[width=1.0\linewidth]{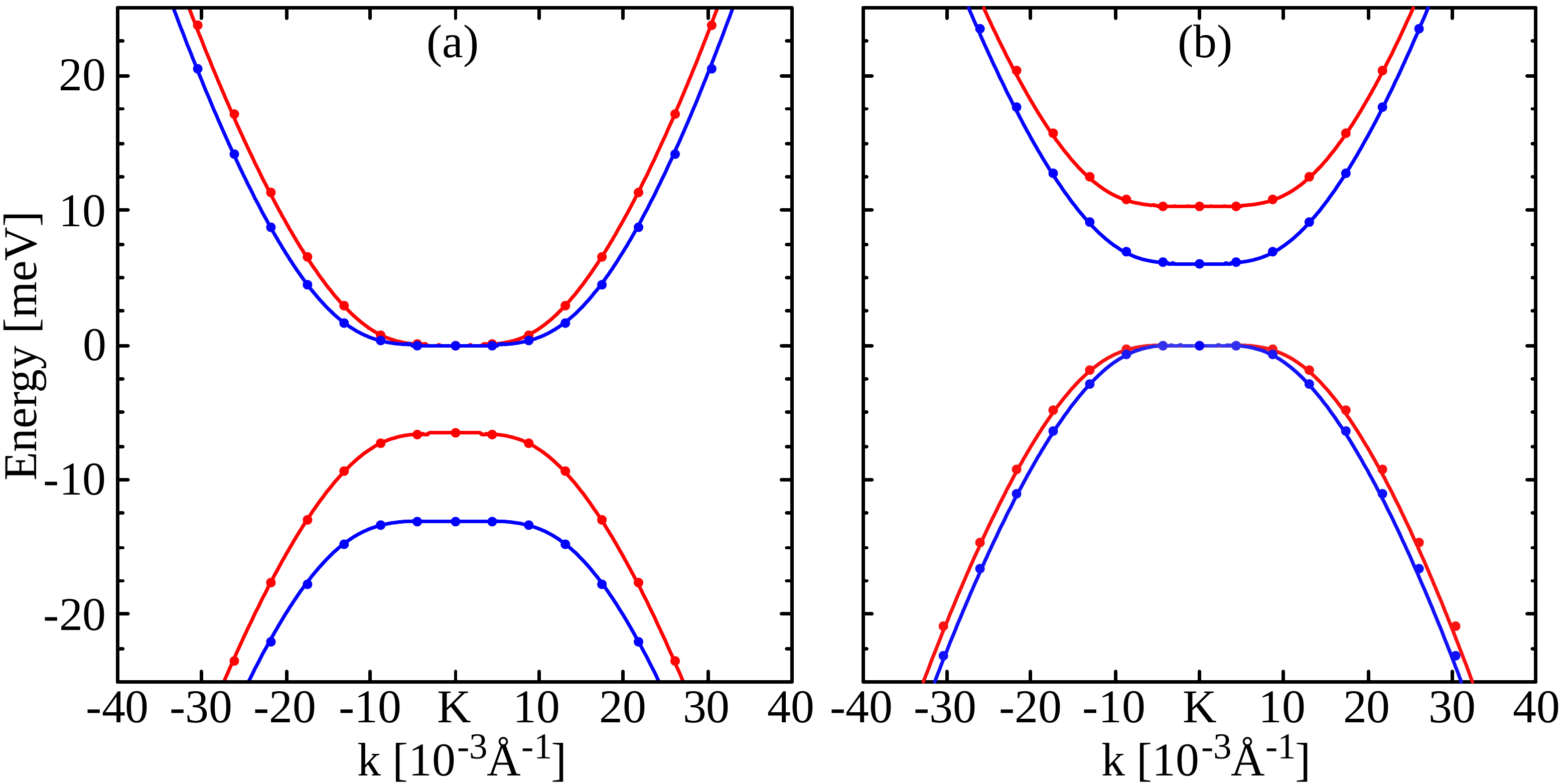}
   	\caption{Low energy band structure of BLG-TMD multilayer system under a gate voltage. Red (blue) lines describe spin down (up) states obtained from tight-binding calculations; circles show results of effective Hamiltonian fit. (a) The neutrality point of BLG is shifted close to the valence bands of the TMD by the gate, resulting in large spin splitting in the valence band of the structure. (b) When the neutrality point is brought close to the conduction band of  the TMD, the overall bandgap is similar ($\simeq 8$ meV here) but spin splitting is larger in the conduction band of the structure. Details of parameters for both panels are given in \cite{supp}. }
   	\label{Fig2}
   \end{figure}

We first consider a gate voltage between BLG and the TMD layer.  This shifts the neutrality point of the unbiased parabolic bands of BLG across the optical gap of the TMD. The resulting bands of the {\em coupled} multilayer are shown in Fig.\ \ref{Fig2}. Conduction and valence bands near the Fermi level develop a gap and visible spin splitting throughout. Gate voltage reversal switches the spin splitting but yields similar bandgaps in the heterostructure ($\simeq 10$ meV) \cite{supp}.
The curvature and detailed topological structure of the states is also controlled by the gate voltage, as we discuss in detail below.
In addition to the overall shift of the neutrality point, a relative potential difference between graphene layers can be applied.  Such difference can be cast as an {\em opposite} or relative voltage between the graphene layers, with respect to the TMD.\@  When present, it increases the overall size of the heterostructure bandgap (similar to the case of isolated BLG \cite{graphene,TI-Bi-G}, while maintaining sizable band inversion and spin splitting effects--see supplement for examples  \cite{supp}.


The knowledge of the dispersions and eigenstates, as well as consideration of the possible symmetry-allowed terms, enables the definition of effective Hamiltonian parameters that fully describe the dispersion and state structure in the respective valleys.
This Hamiltonian respects time reversal symmetry and captures the reduced space symmetries and SO couplings induced in the system. To linear order in momentum (away from each valley), it is given by	
$ \mathcal{H}=\mathcal{H}_{0}+\mathcal{H}_{\Delta}+\mathcal{H}_{S}+\mathcal{H}_{\lambda}+\mathcal{H}_{R}+\mathcal{H}_{V}+\mathcal{H}_{c}$,
	with
	\begin{equation}\label{Eff-H-GGT}
	\begin{split}
	\mathcal{H}_{0}&= v_{F \chi} \left( \tau_z\sigma_{x}p_{x}+\tau_0\sigma_{y}p_{y}\right)s_0\kappa_{0}\;,\\
    \mathcal{H}_{c}&= {\textstyle\frac{1}{2}} t_c\tau_0s_0(\kappa_{x}\sigma_{x}-\kappa_{y}\sigma_{y})\;,\\
	\mathcal{H}_{\Delta}&=\Delta_{\chi} s_0 \sigma_z \tau_0\kappa_{0}\;,\\
	\mathcal{H}_{S}&=S_{\chi} \tau_z\sigma_zs_z\kappa_{0}\;,\\
	\mathcal{H}_{\lambda}&=\lambda_{\chi}\tau_z\sigma_0s_z\kappa_{0}\;,\\
	\mathcal{H}_{R}&= {R}_{\chi} (\vec{\sigma}\times\vec{s}_{\tau})_{\hat{z}}\kappa_{0} \\
                  &\;\;\;\; +\widetilde{R}_{\chi}\left(p_{y}(\vec{\sigma}\times\vec{s}_{\tau})_{\hat{z}}-p_{x}(\vec{s}\cdot\vec{\sigma}_{\tau}
                  )\right) \kappa_0 \; , \\
	\mathcal{H}_{V}&= s_0 \sigma_0 \tau_0(\kappa _{z}V+\kappa_{0}\delta_{\chi}) \;.
	\end{split}
	\end{equation}
Here, $ \sigma_i$, $ \tau_i$, $ s_i$, and $\kappa_i$ are $2\times2$ Pauli matrices (0 is the unit matrix) operating
on various degrees of freedom allowed in the system. $\sigma_i$ acts on the pseudospin sublattice space  (A,B), $ \tau_i $ on the $K, K'$
valley space, $ s_i $ on the spin, and $ \kappa_i$ on the graphene layer index (isospin) of the system, with $\chi=1$ and 2.
We have also defined 
$\vec{s}_{\tau}=(\tau_{0}s_{x},\tau_{z}s_{y},0)$, and $\vec{\sigma}_{\tau}=(3\tau_{z}\sigma_{x},-\tau_{0}\sigma_{y},0)$,
as well as $\widetilde{R}_{\chi}=R_{\chi}a/(4\hbar\sqrt{3})$, with $a=2.46$ \AA.  $\mathcal{H}_{0}$ describes the free particle Hamiltonian of each graphene layer, while $\mathcal{H}_{c}$ describes the coupling in a  Bernal stacking. The TMD leads to the appearance of a staggered potential field described by $\mathcal{H}_{\Delta}$. It also induces SO couplings: the intrinsic $\mathcal{H}_{S}$ preserves in-plane inversion, while the Zeeman-like $\mathcal{H}_{\lambda}$  breaks that symmetry. The reduced symmetry also allows for Rashba couplings described by $\mathcal{H}_{R}$.
Notice these parameters have an implicit dependence on the overall gate shift of the BLG neutrality point with respect to the TMD.  In addition, the
presence of an opposite gate is captured by $\mathcal{H}_{V}$, setting each graphene layer at asymmetric voltages ($\pm V$ when present), with respect to relative shifts ($\delta_{\chi}$).

The effective Hamiltonian in Eq.\ \eqref{Eff-H-GGT} has an explicit dependence on the graphene layer index $\chi$, reflecting differences of the asymmetric coupling between the TMD layer and the two graphene layers. For example, in the system of Fig.\ \ref{Fig2}b, the Zeeman-like SO in the graphene layer closer to the TMD is  $\lambda_1=-3.86$ meV, while in the distant layer
is $\lambda_2=-1.33$ meV.\@
Notice also that Zeeman-like SO couplings $\lambda_\chi$ are dominant, in contrast to others such as the staggered potential and intrinsic SO, $\Delta_{1,2}=-0.24, 0.06$ meV, and $S_{1,2}=-0.26, 1.33$ meV, respectively.
The strong Zeeman-like SO supports the formation of inverted and spin-split bands in the system. The Rashba couplings are also layer asymmetric, with $R_{1,2}= 0.11, 1.1$ meV.
As shown in Fig.\ \ref{Fig2}, the bandgap and inverted curvature of the spectrum around the $K$ valley is fully described by the effective Hamiltonian.  It also captures the full spin, pseudospin, and valley textures of the low energy states, as seen in \cite{supp}.

The BLG-TMD Hamiltonian is characterized by an antiunitary time reversal symmetry and accompanying Kramers degeneracy of states. The system lacks both particle-hole and chiral symmetry, hence it belongs to the AII topological class  \cite{topoclass1},
characterized in two dimensions by a Z$_{2}$ topological invariant.
A continuous tuning of the Hamiltonian parameters may drive topological phase transitions in the system, together with gap closings \cite{topoclass1}.  However,  although gap closing is a necessary condition between different phases, it is not  sufficient, and the change in the Z$_2$ invariant must accompany the topological character change.

\begin{figure}[h]
		\includegraphics[width=1\linewidth]{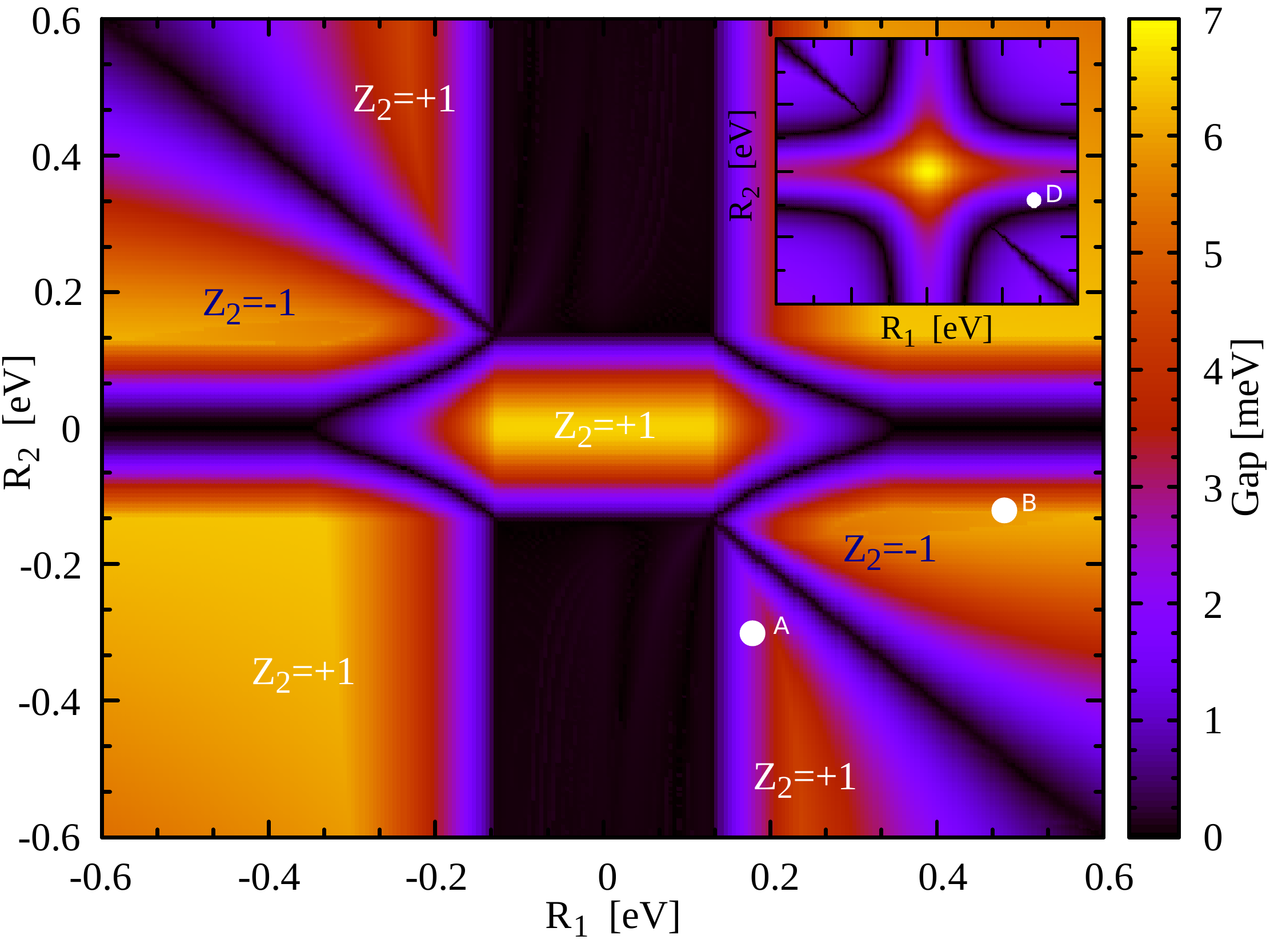}
		\caption{Map of the bulk spectral gap in BLG-TMD heterostructure as function of $R_1$ and $R_2$ for $V=0$.  Notice that only two regions $R_1R_{2}<0$ are topologically nontrivial phases, characterized by index Z$_{2}=-1$ (blue labels).
All other insulating regions are topologically trivial, Z$_{2}=+1$. Inset: Corresponding map of spectral gap for $V=10$ meV.\@ The application of an opposite voltage  drives the entire region into  different {\em trivial} insulating phases, all with Z$_{2}=+1$. The dispersion relation of points A, B, and D, are shown explicitly in Fig.\ \ref{Fig4}. }
		\label{Fig3}
\end{figure}

To analyze the possible phases of the BLG-TMD hybrid, we tune the Rashba couplings $R_{1}, R_{2}$ and opposite voltage $V$, as these parameters may be more easily controllable in experiments.  We first identify the insulating phases of the system described in Fig.\ \ref{Fig2}a, in the absence (Fig.\ \ref{Fig3}) and presence of $V$ (inset of Fig.\ \ref{Fig3}).  Note that the tuning of $R_{1}$ and $R_{2}$ leads to gap closings often away from the $K$-points \cite{Mehdi}.  We emphasize that the maps of spectral gaps obtained both with the tight-binding or effective Hamiltonian are identical over the ranges shown.
	
\begin{figure}[h]
		\centering
	\includegraphics[width=1.0\linewidth]{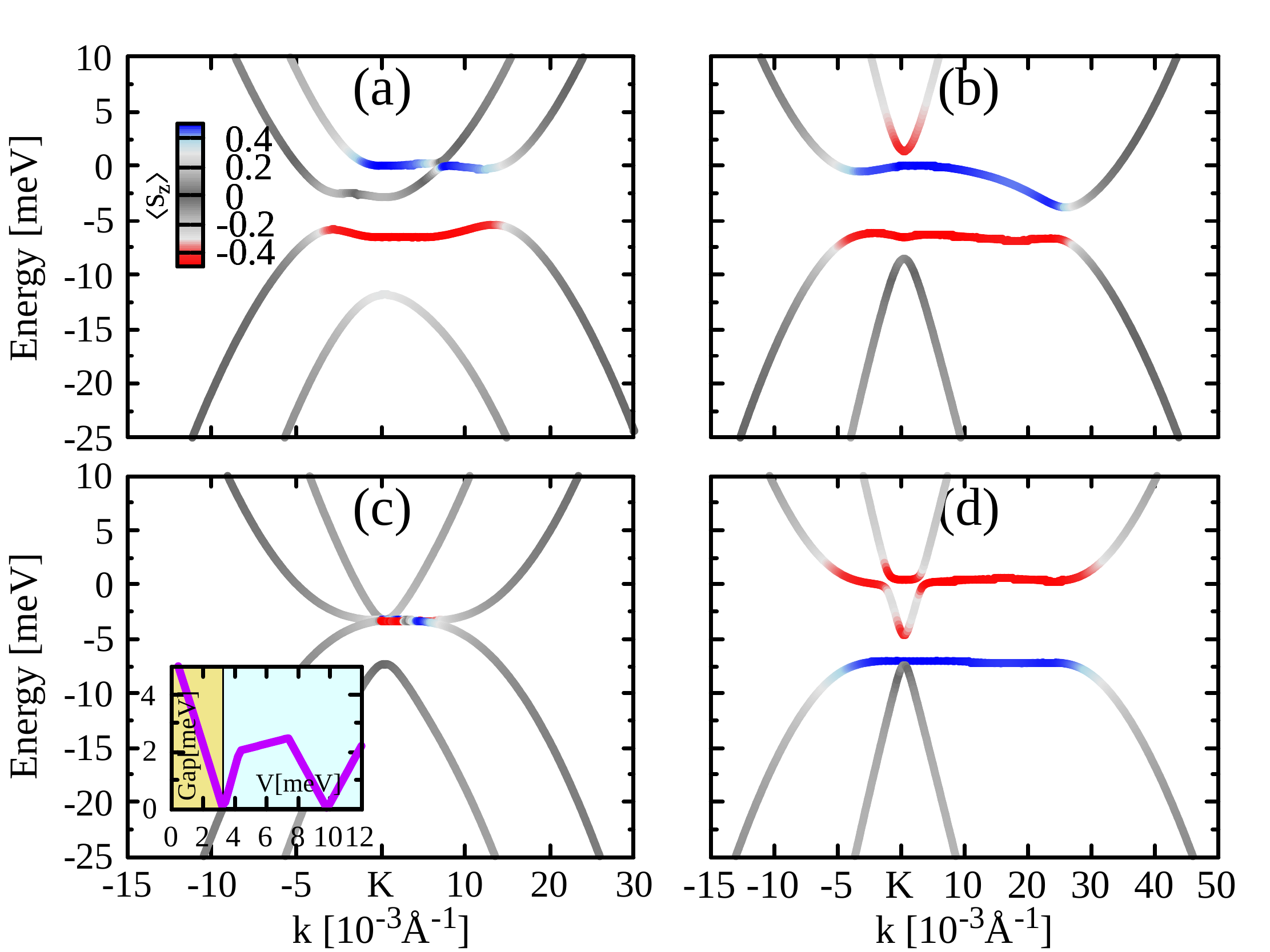}
		\caption{Band structure near the $K$ point of different phases in BLG-TMD.  (a) Shows dispersion relations for trivial insulating phase labeled for $R_{1}, R_{2}$ values indicated by point A in Fig.\ \ref{Fig3}.
(b) Bandstructure for topologically nontrivial insulating phase for point B in Fig.\ \ref{Fig3}. (c) Semimetallic phase separating the nontrivial phase in (b) from the trivial phase (d).  This gapless phase evolves from point B under the application of an opposite voltage $V=3.3$ meV.\@  Inset in (c) shows how gap at B has non-monotonic dependence with opposite voltage, and yet trivial phases for $V > 3.3$ meV.\@
(d) Trivial insulating phase for $R_{1}, R_{2}$ given in (b) but gate voltage $V=10$ meV--point D in Fig.\ \ref{Fig3} inset. Dispersion curves color indicates $s_z$ projection for each state as per legend in panel (a).}
		\label{Fig4}
\end{figure}

As the BLG-TMD system lacks inversion symmetry, the calculation of the Z$_{2}$ invariant cannot be done through analysis of wave function parity at time reversal invariant momenta in the Brillouin zone \cite{TISwithI}.  We use instead two different  methods: the first examines the full non-Abelian adiabatic transport along a time reversal path in the Brillouin zone,
and does not require gauge fixing \cite{Z2-Method,Z2-Method-1,Z2-Method-2}. The second method, implemented in the Z2Pack \cite{Z2Pack}, is based on tracking hybrid Wannier centers for relevant bands.
Both methods yield the same results in this case. The analysis reveals that in the absence of an asymmetric voltage, the system can indeed host topological (Z$_{2}=-1$) and trivial (Z$_{2}=1$) phases, depending on the values of $R_{1}$ and $R_{2}$, as depicted Fig.\ \ref{Fig3}.
Note that topological nontrivial regions require $R_1 R_2 <0$ with minimal magnitudes of $\simeq 100$ meV, and are clearly separated by gap-closings from different trivial phases.
Examples of dispersion relations for topologically distinct regions at points A and B in Fig.\ \ref{Fig3} are shown in Fig.\ \ref{Fig4}, (a) and (b) respectively, and exhibit subtle dispersion and spin texture differences.

Applied gate voltages can also change the topology of the system.
As shown in the inset of Fig.\ \ref{Fig3}, an opposite voltage comparable to the bandgap, $V=10$ meV, makes the system topologically trivial for the entire range of $R_{1}$ and $R_{2}$ shown.  Indeed, as seen in Fig.\ \ref{Fig4}c, a lower gate voltage, $V=3.3$ meV, applied at point B in Fig.\ \ref{Fig3} closes the gap and makes the system semimetallic.  As the voltage increases further, point B transitions to a gapped trivial phase, as indicated by point D in Fig.\ \ref{Fig3} inset, and acquires the dispersion shown in Fig.\ \ref{Fig4}d. Notice that the system described in Fig.\ \ref{Fig2}b has a trivial phase for all opposite voltage values, including $V=0$ \cite{supp}. This highlights the strong SO in the TMD valence bands inherited by the BLG.

The BLG-TMD heterostructure describes an experimentally feasible system with unique tunable topological properties \cite{SOC-G-TMDs}, the result of the reduced symmetries of
the system.  Although a single-layer graphene-TMD can have mass inverted (but topologically trivial) phases, the BLG-TMD system cannot
be seen as two coupled insulating systems, for which the Z$_{2}$ invariant would be the product of the invariants of each subsystem \cite{TISwithI}. The classification here requires consideration of the full structure. The presence of the staggered potential and Zeeman-like SO coupling play a critical role in defining the topological phases in these fascinating experimental systems \footnote{A related structure, where two graphene single layers encapsulate a TMD, do not yield insulating phases.  However, the resulting states do exhibit interesting spin texture--see supplement \cite{supp}.}.


The BLG-TMD hybrid has then gapped regimes for range of experimentally tunable parameters (Rashba coupling and gate potential). The topology of different phases of the system is found to be nontrivial for a range of layer asymmetric values of Rashba interactions.
 Our analysis indicates that a sample can exhibit a quantum spin-Hall phase for Rashba values $R_{1}\approx-100$ meV and $R_{2}\approx 200$ meV.

It is also clear that the experimental observation of the quantum spin-Hall state in these structures relies heavily on the interplay between Rashba SO enhanced in the system, and the gate potential applied. Experiments have reported that the Rashba interaction produced in a BLG-TMD system is typically $\simeq 15$ meV \cite{SOC-G-TMDs}.
This suggests that apart from using different TMD layers \& suitable gates, the experimental realization of a quantum spin-Hall regime would require the enhancement of Rashba SO via alternate or complementary methods, such as heavy atom intercalation and/or deposition, as reported for graphene systems \cite{Berlin,golddep,indiumexpt,hydrogen,colossal,nanoparticles}.

\acknowledgments
We acknowledge support from NSF DMR 1508325 (Ohio),  DMR 1410741 and DMR 1151717 (LSU), and the Saudi Arabian Cultural Mission to the US for a Graduate Scholarship.

\bibliography{cites}
\bibliographystyle{apsrev4-1}

\end{document}